\documentclass[pra,notitlepage]{revtex4-1}
\usepackage{graphicx,color}
\usepackage{dcolumn}
\usepackage{bm}
\usepackage{amsmath,amssymb,mathrsfs,amsthm}
\usepackage{url}
\usepackage{hyperref}

\setcitestyle{sort&compress,numbers}

\newcommand{\C}{\mathbb{C}}

\newcommand{\set}[1]{\mathsf{#1}}

\newcommand{\spc}[1]{\mathcal{#1}}

\newcommand{\Span}{{\mathsf{Span}}}

\def\>{\rangle}
\def\<{\langle}
\def\kk{\>\!\>}
\def\bb{\<\!\<}

\newcommand{\map}[1]{\mathcal{#1}}
\newcommand{\Tr}{\operatorname{Tr}}

\newcommand{\op}[1]{\operatorname{#1}}

\newcommand{\St}{{\mathsf{St}}}

\newcommand{\Chan}{{\mathsf{Chan}}}

\newcommand{\ketvac}{|\text{vac}\>}
\newcommand{\bravac}{\<\text{vac}|}
\newcommand{\vac}{\ketvac\bravac}

\newtheorem{theo}{Theorem}

\newtheorem{lemma}{Lemma}
\newtheorem{prop}{Proposition}

\newtheorem{defi}{Definition}

\def\Proof{{\bf Proof.~}}
\def\qed{$\blacksquare$ \medskip}

\begin{document}
    \title{Quantum communication through  devices with indefinite input-output direction}
    
    \author{Zixuan Liu}
    \affiliation{QICI Quantum Information and Computation Initiative, Department of Computer Science, The University of Hong Kong, Pokfulam Road, Hong Kong}
    \affiliation{HKU-Oxford Joint Laboratory for Quantum Information and Computation}

    \author{Ming Yang}
    \affiliation{Department of Applied Mathematics and Theoretical Physics, Centre for Mathematical Sciences,
University of Cambridge, Wilberforce Road, Cambridge CB3 0WA, United Kingdom}

    \author{Giulio Chiribella}
    \affiliation{QICI Quantum Information and Computation Initiative, Department of Computer Science, The University of Hong Kong, Pokfulam Road, Hong Kong}
    \affiliation{Department of Computer Science, University of Oxford, Wolfson Building, Parks Road, Oxford, UK}
    \affiliation{HKU-Oxford Joint Laboratory for Quantum Information and Computation}
    \affiliation{Perimeter Institute for Theoretical Physics, 31 Caroline Street North, Waterloo,  Ontario, Canada}


    \begin{abstract}
Certain quantum devices, such as half-wave plates and quarter-wave plates in quantum optics,  are bidirectional, meaning that the roles of their input and output ports can be exchanged. 
Bidirectional devices  can be used in  a forward mode and a backward mode, corresponding to two opposite choices of the input-output direction. They can also be used in a coherent superposition of the forward and backward modes,  giving rise to new operations with indefinite  input-output direction.  In this work we explore  the potential of input-output indefiniteness for the transfer of classical and quantum information through noisy channels.  We first formulate a model of  communication from a sender to a receiver via a noisy channel used in indefinite input-output direction. Then, we show that indefiniteness of the input-output direction yields advantages over  standard  communication protocols in which the given noisy channel is used in a fixed input-output direction.  These advantages  range from  a general reduction of noise in bidirectional processes, to heralded noiseless transmission of quantum states, and, in some special cases, to  a complete noise removal.  The noise reduction due to input-output indefiniteness   can be experimentally  demonstrated with current photonic technologies,  providing a way to investigate the operational consequences of exotic scenarios characterised by coherent quantum superpositions of forward-time and backward-time processes. 
  \end{abstract}
  
\nopagebreak
    \maketitle
     
    \section{Introduction}
 Quantum  mechanics offers  new possibilities for  communication technologies, famously including the possibility of secure quantum key distribution \cite{Bennett85, Ekert91}. These possibilities stimulated the study of quantum communication channels and their communication capacities, giving rise to a generalisation of Shannon's information theory known as quantum Shannon theory  \cite{wilde2013quantum}. 
 Traditionally, the communication protocols studied in quantum Shannon theory assumed 
 that the  communication devices were arranged in a classical, well-defined configuration.  Recently, there has been an interest in exploring communication protocols where the configuration of the devices  is controlled by a quantum degree of freedom.  For example, information could travel along different paths from the sender to the receiver, with the choice of path controlled by a quantum system
\cite{aharonov1990superpositions,oi2003interference,gisin2005error,abbott2020communication,chiribella2019quantum,dong2019controlled}. Similarly, the time of transmission could be controlled by a quantum system, giving rise to interference across multiple time bins \cite{kristjansson2021witnessing}.  A further example includes the use of different communication devices in different orders, with the choice of order controlled by a quantum system \cite{Ebler18,Salek18,Chiribella2021Indefinite,procopio2019communication,procopio2020sending,loizeau2020channel,caleffi2020quantum,sazim2021classical,chiribella2021quantum,bhattacharya2021random,guha2022quantum}.  In recent years, a number of communication advantages of quantum control over the communication devices have been experimentally demonstrated \cite{lamoureux2005experimental,goswami2020increasing,guo2020experimental,goswami2020experiments,rubino2021experimental}.
 
 Quantum communication protocols with superpositions  of configurations could potentially be useful in a future  quantum internet \cite{kimble2008quantum}, where quantum messages could be routed coherently along multiple paths.   At the same time, quantum superpositions of alternative configurations have also attracted interest for foundational reasons, including the possibility to explore the operational features  of perspective quantum gravity scenarios  \cite{hardy2007towards,hardy2009quantum,hardy2021time}. More broadly,  quantum communication tasks provide a lens for  understanding the operational consequences of   new foundational  concepts, such as indefinite causal order \cite{chiribella2009beyond,oreshkov2012quantum, araujo2015witnessing,branciard2016witnesses,chiribella2013quantum}. 
 
     \begin{figure}
        \begin{center}         \includegraphics[]{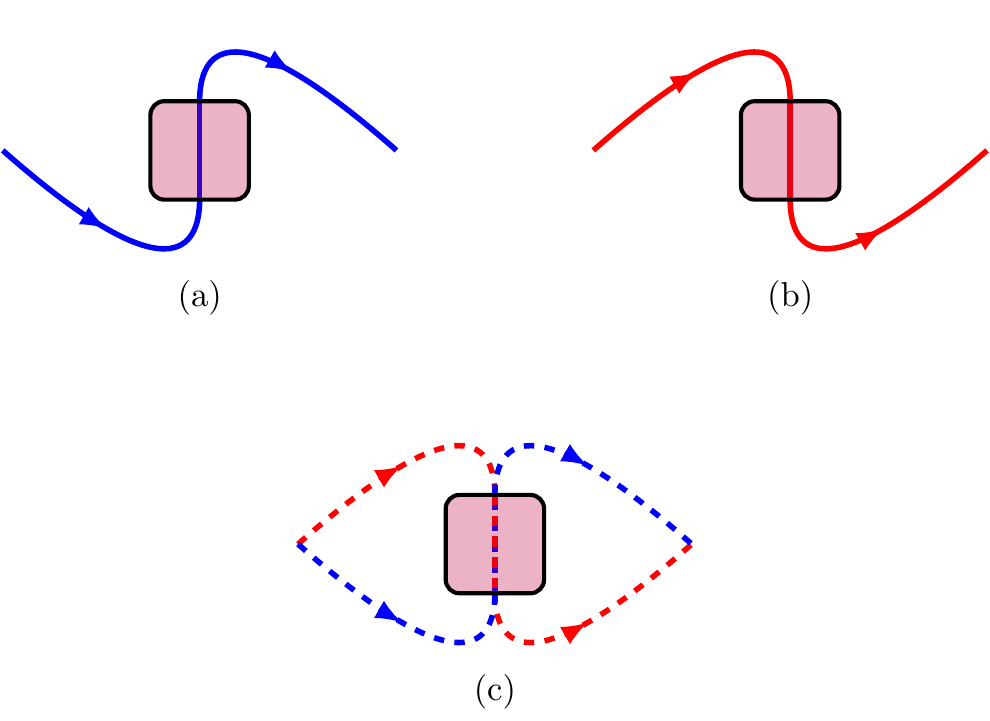}
        \end{center}
        \caption{A bidirectional device, such as a half-wave plate or a quarter-wave plate, can be used in two alternative modes: a forward mode (corresponding to the blue path in subfigure (a)), and a backward mode (corresponding to the red path in subfigure (b)).  Quantum control over the system travelling through the device can be used to create a coherent superposition of the forward and backward mode (subfigure (c)), giving rise to scenarios where the roles of the input and output ports of the device becomes indefinite.}\label{fig:qtf}
    \end{figure}

 A recent example of a new foundational concept is the concept of indefinite time direction  \cite{chiribella2022quantum}. This notion was  originally motivated by  questions about the arrow of time, including whether it is in principle possible to conceive quantum processes that receive their inputs in the future and produce their outputs in the past, and whether it is in principle possible to conceive situations where the time  order between the inputs and the outputs of a quantum process is indefinite.  The study of these foundational questions led to a broader notion   of {\em indefinite input-output direction}, which applies to any scenario involving  bidirectional processes \cite{chiribella2022quantum}, that is,  processes  for which the roles of the input and the output ports can be exchanged.  Examples of bidirectional processes    are the processes induced by half-wave plates, quarter-wave plates, and other optical crystals that rotate the polarisation of  single photons.   Any such  crystal can be traversed in two opposite directions, {\em e.g.} from bottom to top or from top to bottom, as in Fig. \ref{fig:qtf}.  The two opposite ways of traversing the crystal give  rise to  two different quantum processes, related by a symmetry transformation called {\em input-output inversion} \cite{chiribella2022quantum}.  Conventionally, we can call one process the   {\em forward process} and the other process the {\em backward process}.
 Forward and backward processes can be also probed in a coherent quantum superposition, by controlling a photon's trajectory through the crystal, as illustrated in  Fig. \ref{fig:qtf}.  As a result, the direction of the information flow within the crystal can become  indefinite.  
 
 A general mathematical framework for describing the use of a bidirectional device in an indefinite input-output direction was provided in Ref. \cite{chiribella2022quantum}. The framework includes operations that generate a coherent superposition of forward and backward processes, as well as more general operations, therein called operations with indefinite time direction.  The reason for the name is that Ref. \cite{chiribella2022quantum} considered  foundational scenarios  where the two ports  of a bidirectional device are associated to two  specific moments of time $t_1$ and $t_2 > t_1$, so that the ``forward process" would be the process with inputs entering in the past (at time $t_1$) and output exiting in the future (at time $t_2$),  while the ``backward process'' would be the process for which the roles of past and future are exchanged.  Here, instead,  we will use the more neutral term {\em quantum operations indefinite input-output direction}, to stress that the framework applies also to scenarios in which the input and output ports are not associated to specific moments of time.

A first operational consequence of input-output indefiniteness  was shown in Ref. \cite{chiribella2022quantum}, which provided a quantum game in which the player can win with certainty  only if the input-output direction is indefinite. Recently, this game was demonstrated experimentally with photons   \cite{guo2022experimental,stromberg2022experimental}.   Another consequence can be deduced from a related work on the quantum superposition of thermodynamic processes with opposite time arrows \cite{rubino2021quantum}. Besides these works,  however, the operational consequences of indefinite input-output direction are still largely unexplored, especially in comparison to those of indefinite causal order, which have been extensively investigated in  the past decade \cite{chiribella2012perfect,chiribella2013quantum,araujo2014computational,guerin2016exponential,ebler2018enhanced,zhao2020quantum,felce2020quantum,chiribella2021quantum,procopio2015experimental,rubino2017experimental,goswami2018indefinite,wei2019experimental,guo2020experimental,rubino2021experimental,cao2021experimental,nie2022experimental}.   
  
   In this paper, we explore the consequences of input-output indefiniteness for the transmission of   classical and quantum information through noisy channels.  We formulate  a  communication model  in which a sender transmits quantum particles to a receiver through a  bidirectional  device that acts in a coherent superposition of the forward and backward mode, as in   Fig. \ref{fig:qtf}.  
     Specifically, we consider the basic scenario of a single particle travelling through a single quantum device, and we show that the ability to control the direction in which the particle traverses the device  can increase the amount of classical and quantum information reaching the receiver.  For dephasing noise, corresponding e.g. to rotations of a single photon's polarisation about a fixed direction,  we show that a perfect, deterministic transmission of quantum bits becomes possible even if the original dephasing channel was entanglement-breaking and therefore unable to transmit any quantum information.   For more radical types of noise, such as depolarising noise, we show that indefinite input-output direction increases  the classical channel capacity ({\em i.e.} the optimal rate at which the channel can reliably transmit classical bits) and the quantum channel capacity ({\em i.e.} the optimal rate at which the channel can transmit quantum bits).  Notably, both capacities can be non-zero  even in parameter regimes where no information can be transmitted by using the device in a fixed direction.    For example, a completely depolarising qubit channel used in an indefinite input-output direction gives rise to a classical  capacity of $0.3113$ bits per channel use, and even permits a noiseless heralded transmission of qubit states with a success probability of $25\%$.

Our results highlight some similarities, as well as some  differences, between  input-output indefiniteness and  the related notions of indefinite causal order and indefinite trajectories.  Regarding the similarities, the communication advantages observed in all these scenarios are consequences of the ability to coherently control the configuration of noisy channels, as first observed by Gisin {\em et al} \cite{gisin2005error} and recently elaborated in Refs. \cite{abbott2020communication,kristjansson2020resource}.  On the other hand, our results indicate that coherent control over the input-output direction can offer larger advantages.  For example, putting two completely depolarising qubit channels on two alternative paths, and coherently controlling the path gives a classical capacity of at most 0.16 bits per channel use, while  using the two channels in a coherently controlled order yields a capacity of at most 0.049 bits per channel use.  Both values  are strictly smaller than the classical capacity of 0.3113 bits per channel use achievable with a {\em single} depolarising channel used in a coherent superposition of the forward and backward mode. 

Communication protocols exploiting   input-output indefiniteness are also interesting outside the context of quantum communication. They can be regarded as a new type of error correction, boosted by coherent quantum control over the input-output direction of the noisy processes. These protocols have some similarity with dynamical decoupling protocols \cite{viola1998dynamical}, in which an unknown  evolution  is alternated with its inverse in a sequence of time steps. In our protocols, however, the advantages do not come from alternation, but rather from the quantum interference between a forward process and the corresponding backward process. 
As we will see later in the paper, the working principle of our protocols  appears to be the ability to distinguish between errors represented by symmetric matrices and errors represented by anti-symmetric matrices, which  in particular allows one to correct  Pauli $Y$ errors separately from the other types of single-qubit errors.

 The rest of the paper is organised as follows. In Section \ref{sec:model} we introduce a communication model that uses  bidirectional devices in a coherent superposition of the forward and  backward directions.   In Section \ref{sec:transpose} we apply our communication model to the special case of noisy channels that are invariant under transposition, which includes all the examples studied in the rest of the paper.     The examples are provided in Section \ref{sec:advantages}, where we show  that  input-output  indefiniteness can boost the capacity of depolarising and dephasing channels.  Finally, the conclusions are drawn in   Section \ref{sec:summary}.

    \section{Communication model with superposition of input-output directions}\label{sec:model}
    
        \subsection{Input-output inversion and superposition of input-output directions}
    \label{subsec:indefinitetime}
    Let us start by reviewing the notion of indefinite input-output direction introduced in Ref. \cite{chiribella2022quantum}. This notion applies to bidirectional quantum processes, that is, quantum processes for which roles of the input and output ports are exchangeable.  For example, the transmission of a single photon through a half-wave plate is a bidirectional process, because the exchange of the  entrance point of the photon  with its exit point gives rise to a valid single-photon evolution.
     
     
   The set of bidirectional quantum processes on a given quantum system was characterised in Ref. \cite{chiribella2022quantum}. There, it was shown that a process is bidirectional if and only if it is described by a bistochastic channel \cite{landau1993Birkhoff,mendl2009unital}, that is, a completely positive linear map  $\map C$ with a Kraus representation 
       $\map C(\cdot) = \sum_{i=1}^r C_i \cdot C_i^\dagger$ where    
the Kraus operators $\{ C_i \}_{i=1}^r$ satisfy the normalisation condition $\sum_{i=1}^r C_i^\dagger C_i =  \sum_{i=1}^r C_i C_i^\dag   =  I$,  $I$ being the identity matrix on the system's Hilbert space. 

A given bidirectional device is associated to a forward channel $\map C$  (describing the transformation of the system's density matrix in a given direction, conventionally regarded as the ``forward'' one)  and to a backward channel $\Theta (\map C)$  (describing the transformation of the system's density matrix in the opposite direction). The map $\Theta$ transforming the forward channel into the backward channel is called an {\em input-output inversion}.  Ref. \cite{chiribella2022quantum} showed that the possible input-output inversions are of the form $\Theta  (\map C)  (\cdot)   =  \sum_i  \theta  (C_i)     \cdot  \theta  (C_i^\dag)$, where the matrices  $\theta  (C_i)$ are either unitarily equivalent to the transpose, namely  
\begin{align}\label{thetatranspose}
\theta  (C_i)   = U C_i^T  U^\dag \,    \qquad \forall i \in   \{1,\dots, r\}
    \end{align}
where $U$ is a fixed  unitary matrix and $C_i^T$ is the transpose of $C_i$ in a fixed basis,   
or unitarily equivalent to the adjoint, namely
\begin{align}\label{thetaadjoint}
\theta  (C_i)   = U C_i^\dag  U^\dag \, ,   \qquad  \forall i\in  \{1,\dots,  r\} \,,   
    \end{align}
where $C_i^\dag   $ is the adjoint (conjugate transpose) of $C_i$. 
A fundamental difference between the transpose and the adjoint is that the transpose can in principle be applied locally to part of a bipartite quantum channel, whereas the adjoint cannot, because it would generally give rise to maps that are not completely positive \cite{chiribella2022quantum}.   Technically, the difference is that the transpose is an admissible transformation of quantum channels  (an admissible quantum supermap, in the language of  \cite{chiribella2008transforming}), while the adjoint is not.  In the following, we will focus our attention to bidirectional devices for which the input-output inversion is unitarily equivalent to the transpose. Example  are photonic devices, such as half-wave plates and other polarisation rotators, or Stern-Gerlach devices.

Another example of input-output inversion is  time reversal   \cite{wigner1959group,messiah1965quantum}. An extended  discussion of this example is provided in  \cite{chiribella2022quantum}, and here we only sketch the main ideas for the reader's convenience.  The standard approach to time reversal, dating back to Winger \cite{wigner1959group},   is to define an anti-unitary operation $T$ acting on quantum states.  The operator $T$ is generally  taken to represent the change of description  due to the inversion of time coordinate $t\mapsto -t$.  Now, the  time reversal of quantum states implicitly defines  a time reversal of unitary evolutions: if a unitary evolution $U_{t_0, t_1}$ transforms the state $|\psi\>$ at time $t_1$ into the state  $U_{t_1,t_2}  |\psi\>$ at time $t_2  >t_1$, then the time-reversed evolution  should transform the state $ T U_{t_1,t_2}|\psi\> $ at time $-t_2$ into the state $T|\psi\>$ at time $-t_1$. Since this condition should hold for every state $|\psi\>$,  setting $|\varphi\>  :=  T U_{t_1,t_2}|\psi\>$ we obtain that  the time-reversed evolution, denoted by $\theta(U_{t_1, t_2}) $, must transform the state $|\varphi\>$ into the state $T   U_{t_1,t_2}^\dag   T^{-1}   |\varphi\>$.  Hence, the time-reversed evolution is $\theta(U_{t_1, t_2})   =T  U_{t_1,t_2}^\dag  T^{-1}$.  This expression is consistent with  Eq. (\ref{thetatranspose}):   by decomposing anti-unitary operator  as $T  =  U  K$, where $U$ is a unitary operator and $K$ is the complex conjugation, we obtain the expression  $\theta(U_{t_1, t_2})   =U \, U_{t_1,t_2}^T  \, U^\dag$, meaning that  time reversal of the unitary evolution $U_{t_1,t_2}$ is unitarily equivalent to its transpose $U_{t_1,t_2}^T$.  The above argument can be  extended from unitary channels to general bistochastic channels by using the fact that bistochastic channels are linear combinations of unitary channels (see \cite{chiribella2022quantum} for more details).  In summary, the canonical time reversal in quantum theory is associated to an input-output inversion that is unitarily equivalent to the transpose.  Furthermore, the combination of time reversal  with other  symmetries, such as charge conjugation $C$ and  parity inversion $P$, gives rise to other examples of input-output inversions, such as those corresponding to the symmetries $CT$, $PT$, and $CPT$. More generally, an  input-output inversion  involves the exchange of two ports of a quantum device, without necessarily associating the two ports to two specific moments of time $t_1$ and $t_2$. 

Traditionally, the roles of the input and output ports of quantum devices  have been taken to be fixed. 
In general, however, the user  is free to choose  which port serves as the input and which one serves as the output, and the choice can even be controlled by the state of a quantum system.  
The ability to control the input-output direction of the information flow within a given device is described by a higher order transformation, called the {\em quantum time flip} \cite{chiribella2022quantum}. The quantum time flip takes as input a bistochastic  quantum channel $\map C$,    and produces as output another bistochastic  quantum channel $\map F  (\map C)$, acting on the original quantum system  (hereafter called the {\em target}) and on a {\em control} qubit, which determines the input-output direction.  The channel $\map F (\map C)$ can be specified via its Kraus representation, given by  $\map F   (\map C)  (\cdot)  =  \sum_i  F_i  \cdot F_i^\dag$ with 
    \begin{equation}
        \label{eq:qtf}
        F_i = C_i \otimes |0\>\<0| + C_i^T \otimes |1\>\<1| \quad i \in \{ 1, \cdots, r \} \, .
    \end{equation}
    If the control system is set to $|0\>$, then the action of the channel $\map F (\map C)$ on  the target system  is given by  the forward  channel $\map C$ (Fig. \ref{fig:qtf}(a)); if instead the control system is set to $|1\>$, then the action on the target  is given by the backward channel   $\map C^T$ (Fig. \ref{fig:qtf}(b)). Indefiniteness of the input-output direction arises when the control system is set to a coherent superposition of $|0\>$ and $|1\>$, such as {\em e.g.} the maximally coherent state $|+\rangle  =  (|0\>  + |1\>)/\sqrt 2$. In this case,  the channel $\map F (\map C)$ can be regarded as a coherent superposition of the forward and backward channels $\map C$ and $\map C^T$, in the sense of Ref. \cite{aharonov1990superpositions,aaberg2004subspace,aaberg2004operations,oi2003interference,oi2003interference,zhou2011adding,soeda2013limitations,araujo2014quantum,friis2014implementing,gavorova2020topological,thompson2018quantum,dong2019controlled,gisin2005error,abbott2020communication,chiribella2019quantum,kristjansson2021witnessing,kristjansson2020resource,lamoureux2005experimental,rubino2021experimental} (Fig. \ref{fig:qtf}(c)).

    \subsection{Communication through a single bistochastic channel in an indefinite input-output direction}

       \begin{figure}
        \centering
        \includegraphics[]{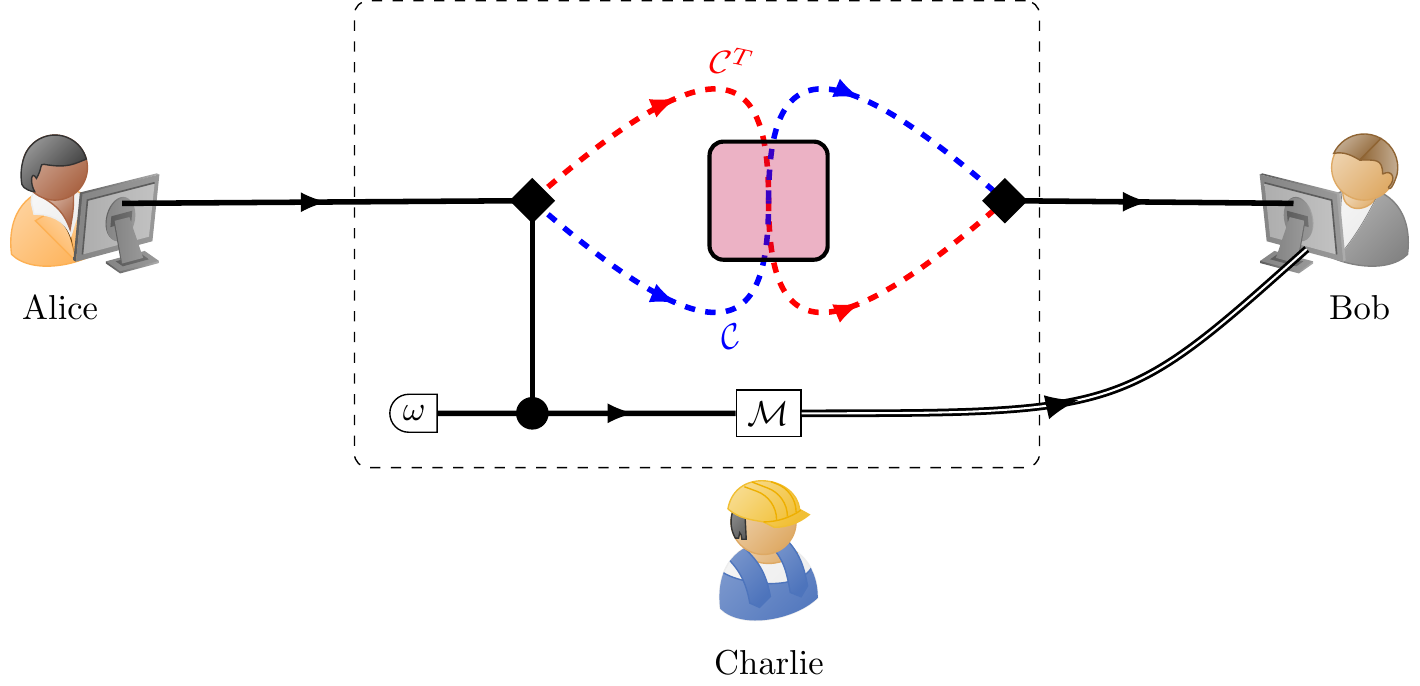}
        \caption{  \textit{Communication model using channels in an indefinite input-output direction}. A sender, Alice, transmits quantum states to a receiver, Bob, with the assistance of a third party Charlie.  The overall channel connecting Alice to Bob  is represented by the dashed black box in the figure, with the black solid line on the left representing Alice's input to the channel, the black solid line on the right representing the channel's output received by Bob, and the double black line on the left representing the transmission of  classical information  from Charlie to Bob.   Inside the box, Alice's input is routed on two possible paths, traversing a device (pink box) depending on the state of a control system (black line below the pink box).   One path (dashed blue line)  traverses the device  in the forward direction (from bottom to top) and is associated to the channel $\map C$. The other path (dashed red line)  traverses  the device in  the backward direction (from top to bottom) and is associated to the channel  $\map C^T$.    The control system is initialised by Charlie in a fixed state $\omega$, independently on the quantum states used by the sender to transmit information to the receiver.  At the end, Charlie performs a measurement $\map M$   on the control system and communicates his measurement outcome to Bob via a classical communication channel. }
        \label{fig:model}
    \end{figure}
     
We now  develop a model of quantum communication using bidirectional devices in an indefinite input-output direction.  The model is 
illustrated  in Fig. \ref{fig:model}.  A sender (Alice)  wants to transmit a quantum state $\rho$  to a receiver (Bob). To this purpose, Alice uses a bidirectional device that acts as  channel $\map C$ when traversed in a given direction and as  channel $\map C^T$ when traversed in the opposite direction.    A third party, Charlie,  controls the direction in which Alice's  particle  traverses  the device, thereby realising the quantum time flip $\map F$ operation described in the previous subsection.  At the beginning of the protocol,  Charlie initialises the control qubit in   a quantum state $\omega$. Then, the target and control, initially in the product state $\rho\otimes \omega$,   undergo the joint evolution described by the quantum channel  $\map F (\map C)$, thereby evolving into the state $[\map F(\map C)]  (\rho\otimes \omega)$.  At this point,  Charlie  performs a measurement  $\map M$ on  the control system and communicates the outcome to Bob via a classical communication channel. 
Finally, Bob  tries to retrieve Alice's input state $\rho$, using the classical information provided by Charlie.

The above communication model can be applied to different communication tasks, including the communication of classical data (in which case the input state $\rho$ is chosen from a set of states $\{  \rho_x\}_{x\in\set X}$ used to encoded a message $x$ from some finite alphabet $\set X$) and the communication of quantum data (in which case the input state $\rho$ is an arbitrary density matrix with support contained in a suitable subspace).   In turn, the communication of quantum data can be used for a number of applications, including for example quantum key distribution  \cite{Bennett85,Ekert91} or quantum secure direct communication \cite{long2002theoretically,deng2003two,long2007quantum}

Let us now  analyze explicitly the steps of our communication protocol. At the beginning, the target system is in the state $\rho$ and the control system is in the state $\omega$.   Then, the two states are sent through the channel $\map F(\map C)$.  The effective evolution from Alice's input state $\rho$ to the joint output state of the target and control is given by a channel $\map F_\omega (\rho)$, defined as
\begin{align}
\nonumber \left[ \map F_{\omega}  (\map C)\right](\rho)
        &:= [\map F(\map C)]\, (\rho \otimes \omega)\\
 \nonumber         & = \sum_{i=1}^r\Big\{    \frac{ C_i  +  C_i^T}2  \,  ~\rho ~ \frac{ C_i^\dag  +  \overline C_i}2   \otimes \omega   +    \frac{ C_i  -  C_i^T}2  \,  ~\rho ~ \frac{ C_i^\dag  -  \overline C_i}2   \otimes Z\omega Z  \\
   &   \qquad   +    \frac{C_i  +  C_i^T}2  \,  ~\rho ~ \frac{ C_i^\dag  -  \overline C_i}2   \otimes \omega   Z  +    \frac{ C_i  -  C_i^T}2  \,  ~\rho ~ \frac{ C_i^\dag  +  \overline C_i}2   \otimes Z\omega   \Big\} \, ,  \label{flippedC}
\end{align}
where $(C_i)_{i=1}^r$  is a Kraus representation of the channel $\map C$,  $\overline  C_i  :=   \left( C_i^T \right)^\dag$ is the complex conjugate of the matrix $C_i$, and $Z  :=  |0\>\<0|  -  |1\>\<1|$ is the Pauli $z$ matrix.    Here, the second equality  follows by substituting  the expressions $|0\>\<0|   =  (I+Z)/2$  and $|1\>\<1|  =  (I-Z)/2$ into Eq. (\ref{eq:qtf}).     The channel $\map F_\omega(\map C)$ describes the use of a bidirectional device in a combination of the forward and backward directions determined by the quantum state $\omega$.   
In the following, we will call $\map F_\omega (\map C)$ the {\em flipped channel}.

The next step in the protocol is Charlie's measurement on the control qubit.   The measurement can be  described by a quantum-to-classical channel $\map M$, of the form  
\begin{align}
\map M  (\rho)    =  \sum_{j=1}^N  \,  |j\>\<j|  \,\Tr [ P_j \rho ] \,,
\end{align}
where $(P_j)_{j=1}^N$ is a positive operator-valued measure (POVM) and $\{|j\>\}_{j=1}^N$ are orthogonal states of a classical system, on which the measurement outcome is recorded.  

After Charlie's measurement, the target system and the classical system end up in the quantum-classical state 
\begin{align}
\rho_{\rm target, classical}  = (\map I_{\rm target} \otimes \map M)  [\map F (\map C) ] \,    (  \rho \otimes \omega)    = (\map I_{\rm target} \otimes \map M)  [\map F_\omega (\map C) ] \,    (  \rho) \, , 
\end{align}
where $\map I_{\rm target}$ is the identity map acting on the target system. 
    Finally,  Charlie communicates to Bob the outcome of his measurement via a classical communication channel. 
    Hence, the classical system at the output of channel $\map M$ is eventually in Bob's laboratory.  

Overall, the flow of information from Alice to Bob is  described by the effective channel 
\begin{align}\label{Ceff}
\map C_{\rm eff}   :  =      (  \map I_{\rm target}  \otimes \map M)   \map F_\omega   (\map C)   .      
\end{align}
The goal of this paper is the study of the classical and quantum capacities of  the effective channel $\map C_{\rm eff}$, optimised over all possible state preparations $\omega$ and over all possible measurements $\map M$.

\subsection{Relations with other communication models in the literature}  

To better understand our communication model, it is useful to consider its  similarities and differences with  other models considered in the literature.  First, it is important to observe  that our model is different from two-way communication models in which information travels simultaneously from Alice to Bob and from Bob to Alice   \cite{del2018two,hsu2020carrying, del2020coherence}, or from related models of multiple access quantum communication channels \cite{zhang2022building,horvat2021quantum}.   Unlike these models, our model does not exhibit any indefiniteness in the roles of sender and receiver, but only in the direction in which the sender's messages traverse  a given quantum device. 

Second, it is important to clarify the role of the third party, Charlie. In our model, one  can think of Charlie as a communication provider, who assists Alice and Bob, without interfering  with Alice's choice of input states.  Accordingly, Charlie initialises the control system  in  fixed quantum state $\omega$, independent of the quantum states sent by Alice. This setting is broadly assumed in other communication models based on superposition of channels \cite{kristjansson2020resource, abbott2020communication, chiribella2019quantum, Chiribella2021Indefinite}. 

It is  also worth noting that Charlie's assistance is limited to the transmission of classical information. 
This constraint is similar to the constraint adopted in the study of  quantum communication with the assistance of classical information from the environment \cite{gregoratti2003quantum,smolin2005entanglement}, with the only difference that in our case we  do not assume assistance from the whole environment of the channel between Alice and Bob, but only from a two-dimensional system  that controls the direction of the information flow within a given device.

On a more formal level, one can regard our communication model as part of a resource theory  \cite{Chitambar2019quantum}. More specifically,    our model is an instance of a resource theory of quantum communication  \cite{kristjansson2020resource}, in which communication devices are   
assembled by the communication provider to build an effective communication channel connecting the sender to the receiver. In this framework, the operations performed by the communication provider are described by quantum supermaps \cite{chiribella2013quantum,chiribella2008transforming,chiribella2009theoretical,bisio2019theoretical}, that is, higher order operations that take in input the quantum channels associated to the initial devices, and produce in output a new quantum channel connecting  the sender to the receiver.

The supermaps implementable by the communication provider are called  {\em free},  in the sense that they are regarded as achievable in the resource theory under consideration. Often, free operations are motivated by practical constraints arising in real-world applications.   Note, however, that in general this needs not be the case:  the resource-theoretic framework is useful not only as a way to analyze practical constraints, but also as a theoretical lens for understanding the power of certain sets of operations, conventionally referred to  as ``free'' even though they are not necessarily  easy to implement. 

Different resource theories of communication  correspond to different choices of free supermaps.   For example, standard quantum Shannon theory can be formulated as a resource theory of communication where the  free supermaps combine the original devices in a well-defined configuration, typically in parallel and without the addition of any other channel on the side \cite{kristjansson2020resource}. Various extension of standard quantum Shannon theory can be obtained by enlarging the set of free operations, {\em e.g.} by including the ability to connect quantum channels in a coherent superposition of orders \cite{Ebler18,Salek18,Chiribella2021Indefinite,procopio2019communication,procopio2020sending,loizeau2020channel,bhattacharya2021random} or to coherently control which quantum channel is used to transmit information \cite{abbott2020communication,chiribella2019quantum}. Communication advantages over standard quantum Shannon theory can then be understood as a result of the enlargement of the set of free supermaps, corresponding to more general ways to assemble the initial devices into an effective channel connecting the sender to the receiver.

The communication model considered in this paper corresponds to a new  way to enlarge the set of free supermaps, including the possibility of using bidirectional devices in a coherent superposition of two  alternative directions.
As we will see later in the paper, this enlargement provides some interesting advantages over the standard model of quantum Shannon theory.    On the other hand,  the model still satisfies a constraint     put forward in Ref. \cite{kristjansson2020resource}:  free operations should not allow Alice and Bob to communicate independently of the initial devices used  by Charlie  to set up the communication link  between them.   The proof that our communication model  satisfies this constraint is provided in Appendix \ref{app:proof}.

\section{Communication through  transposition invariant channels}\label{sec:transpose}
All the concrete examples of noisy channels studied in this paper  are invariant under transposition,  namely  $\map C^T  =   \map C$.  For this class of channels we now derive  Charlie's optimal state preparation $\omega$ and optimal measurement $\map M$.

Let us  start from a characterisation of the transposition invariant channels:

\begin{prop}\label{prop:invariant}
The following are equivalent: 
\begin{enumerate}
  \item  $\map C^T  =  \map C$.
      \item There exists a Kraus representation $\map C  (\rho)  =  \sum_i  C_i  \rho C_i^\dag$ in which each Kraus operator $C_i$ is either symmetric ($C_i^T  = C_i$) or anti-symmetric ($C_i^T  =  - C_i$). 
\end{enumerate}
\end{prop}
The proof is provided in Appendix \ref{app:invariant}.    

For a transposition invariant channel, the forward and backward processes coincide.  Nevertheless, quantum interference between these two processes  gives rise to  non-trivial effects.  This fact can be seen from Eq. (\ref{flippedC}), which in the case of transposition invariant channels becomes 
\begin{align}\label{simplified}
   \big[ \map F_\omega  (\map C)\big]  \,   (\rho)     =  &  \sum_{i \in  \set S_{\rm sym} }    C_i\rho  C_i^\dag   \otimes \omega   +   \sum_{i \in  \set S_{\rm anti}  } C_i\rho  C_i^\dag   \otimes Z\omega Z  \, , 
\end{align}
where $\set S_{\rm sym}  :  =  \{  i~|~  C_i^T=  C_i  \} $ is the set of indices corresponding to symmetric Kraus operators and  $\set S_{\rm anti}  :  =  \{  i~|~  C_i^T=  -C_i  \} $ is the set of indices corresponding to antisymmetric Kraus operators.

Eq. (\ref{simplified})   tells us that the interference between the two input-output directions separates the symmetric  and anti-symmetric Kraus operators of the channel $\map C$,   correlating the symmetric operators with the state $\omega$, and the anti-symmetric operators with the state $Z\omega Z$.   In particular, if the control state $\omega$ is initially set to the maximally coherent state $\omega =  |+\>\<+| $,  then the output state is 
\begin{align}\label{simplified+}
   \big[ \map F_{|+\>\<+|}  (\map C)\big]  \,   (\rho)     =    \sum_{i \in  \set S_{\rm sym} }    C_i\rho  C_i^\dag   \otimes |+\>\<+|   +   \sum_{i \in  \set S_{\rm anti}  } C_i\rho  C_i^\dag   \otimes |-\>\<-|  \, , 
\end{align}
having used the notation $|\pm  \>:  =  (  |0\>  +\pm  |1\>)/\sqrt 2$.
 Eq. (\ref{simplified+}) shows that the symmetric and anti-symmetric part of channel $\map C$ can be perfectly separated by measuring the control on the orthonormal basis $\{|+\> , |-\>\}$.
 

The above observations provide the optimal strategy for Charlie to assist Alice and Bob:
\begin{prop}\label{prop:optimalprotocol}
 Let $\map C$ be a transposition invariant channel.  For  every communication task in the model of Fig. \ref{fig:model},  one can assume without loss of generality that 
\begin{enumerate}
    \item the control system is initialised in the maximally coherent state $\omega=  |+\>\<+|$, and
    \item the control system is measured on the Fourier basis 
     $\{|+\>,  |-\>\}$, corresponding to the quantum-to-classical channel 
     $\map M_{\rm Fourier} (\rho)  =  |+\>\<+|  \,  \<+|  \rho  |+\>  +  |-\>\<-| \,  \<-|\rho|-\>$. 
\end{enumerate}
\end{prop}

\Proof   The proof is a straightforward consequence of Eqs. (\ref{simplified})  and (\ref{simplified+}). 

Item 1 follows from the fact that Charlie can always obtain  the output state in Eq. (\ref{simplified}) from the output state in Eq. (\ref{simplified+}) by applying a local operation on the control system: Charlie only needs to  measure the control system in the Fourier basis $\{  |+\>,  |-\>\}$ and to reset  the control system  to either the state $\omega$ or to the state $Z\omega Z$, depending on whether the measurement outcome is $+$ or $-$, respectively.  

Item 2 follows from the fact that, for the optimal input state $\omega  =  |+\>\<+|$,  the output state $[\map F_\omega (\map C)] (\rho) $ is invariant under the action of the classical-to-quantum channel $\map M_{\rm Fourier}$.  Hence, transmitting the  outcome of the Fourier basis measurement to Bob is equivalent to transmitting the state of the control qubit.  Any  other measurement $\map M$ that Charlie could have performed instead of the Fourier measurement can then be performed by Bob as part of his decoding operations.  
 \qed 
 
Thanks to Proposition \ref{prop:optimalprotocol}, we can assume without loss of generality that the control system is initialised in the maximally coherent state $\omega  =  |+\>\<+|$ and measured in the Fourier basis.  In this case, the effective channel $\map C_{\rm eff}$ in Eq. (\ref{Ceff}) is simply given by  
\begin{align}\label{Ceffshort}
\map C_{\rm eff}   :  =       \map F_{|+\>\<+|}   (\map C)\, .   
\end{align}

In the next section, we will study the classical and quantum communication capacities of the channel $\map F_{|+\>\<+|}  (\map C)$ in the case of qubit depolarising and dephasing channels.   Before doing that, we now review the basic notions on classical and quantum channel capacities, and the relation of the latter with the transmission of private classical information. Readers who are already familiar with these notions can skip directly  to the next section.

The classical capacity of a given noisy channel $\map N$ is operationally defined as the maximum number of  classical bits one can reliably transmit  per use of the channel $\map N$ in the asymptotic limit of many parallel  uses.   The relevant quantity for the calculation of the classical capacity is the Holevo information of the channel $\map N$  \cite{Holevo98}, defined as 
\begin{align}
\chi(\map N):= \max_{ (p_x,\rho_x)_{x\in\set X}}    H  \left(  \map N \left(   \sum_{x\in \set X} p_x\, \rho_x  \right)  \,  \right)  -  \sum_{x\in\set X}  p_x \,H \big(  \map N(\rho_x) \big )\,,
\end{align}
where $H  (\rho)  =  -\Tr[\rho \log \rho]$ is the von Neumann entropy of an arbitrary state $\rho$, and the maximisation runs over all possible ensembles $\{  p_x, \rho_x\}_{x\in\set X}$ consisting of density matrices   $\rho_x$ given with probabilities $p_x$. 

As a consequence of the Holevo-Schumacher-Westmoreland theorem \cite{Holevo98,Schumacher97}, the classical capacity is  given by  
\begin{align}
C(\map N)   :=\lim_{k\to \infty}   \frac{  \chi  (\map N^{\otimes k})}{k} \, .   
\end{align}
  Since the Holevo quantity is generally superadditive under tensor products, one generally has the bound 
  \begin{align}\label{boundC}
  C(\map N)  \ge   \chi (\map N)  \, ,  
  \end{align}
valid for arbitrary quantum channels $\map N$.  

Let us consider now the quantum capacity, operationally defined as the maximum number of quantum bits one can reliably transmit per channel use in the asymptotic limit of many parallel uses of the channel $\map N$.  The relevant quantity for the calculation of the quantum capacity is the  coherent information of the channel $\map N$, defined as  \cite{Schumacher96} 
\begin{align}\label{cohinfo}
I_{\rm c}   (\map N)  :  =  \max_{\rho  \in  \St  (A)}    I_{\rm c}  (\map N,  \rho)  \qquad  {\rm with} \qquad     I_{\rm c}  (\map N,  \rho)    :  =    H   \big( \map N (\rho)\big)-   H    \big(   \,  (\map N\otimes \map I)   (\Psi_\rho)  \, \big) \, ,
\end{align}
where $\St (A)$ is the set of  all possible density matrices of system $A$, the input system of channel $\map N$, and $\Psi_\rho\in \St (A\otimes R)$ is an arbitrary purification of the state $\rho$, namely a pure state of a composite system $A \otimes R$  (with a suitable reference system $R$) such that $\rho  =  \Tr_R  [\Psi_\rho]$.  

The Lloyd-Shor-Devetak theorem \cite{Lloyd97,Shor02,Devetak05} shows that the quantum capacity is  given by 
\begin{align}
Q  (\map N)=   \lim_{n\to \infty}  \frac{I_{\rm c}(\map N^{\otimes n})}n 
\end{align}
 Since in general the coherent information is superadditive under tensor products, one has the lower bound 
 \begin{align}\label{boundQ}
 Q(\map N)  \ge  I_{\rm c}  (\map N) \, ,
 \end{align}
valid for every quantum channel $\map N$.  

In turn, the quantum capacity is a lower bound to the private capacity $P(\map N)$ of the channel $\map N$, defined as the maximum number of classical bits that the sender can privately transmit to the receiver  per channel use in the asymptotic limit of many parallel uses.  The relevant quantity for the calculation of the private capacity is the private information of the channel $\map N$, defined as  \cite{Devetak05}
\begin{align}
I_{\rm p}  (\map N)   =   \max_{ (p_x,\rho_x)_{x\in\set X}}   \left\{ \,I_{\rm c}  \left(  \map N  ,  \sum_{x\in\set X} p_x\,  \rho_x\right)   -  \sum_{x\in\set X}  p_x\,  I_{\rm c}  (\map N,  \rho_x) \right\} \, ,
\end{align}
where the maximisation is over all possible input ensembles $(p_x,\rho_x)_{x\in\set X}$.  

Explicitly, the private capacity is given by \cite{Devetak05}
\begin{align}
P  (\map N)=   \lim_{n\to \infty}  \frac{I_{\rm p}(\map N^{\otimes n})}n \, ,
\end{align}
and one has the bound 
\begin{align}\label{privatebound}
P(\map N)  \ge Q(\map N)  \,. 
\end{align}
The intuitive reason for the bound  (\ref{privatebound}) is that the  channel $\map N$   can be used to send private classical messages by encoding them into quantum states, as it is done, for example, in  the protocols of quantum secure direct communication  \cite{long2002theoretically,deng2003two,long2007quantum}.  Hence, the rate  at which quantum states can be transmitted reliably provides a lower bound to the rate at which classical messages can be transmitted reliably {\em and privately}.  

In the following, we will use the lower bounds (\ref{boundC}) and (\ref{boundQ})  to estimate the classical and quantum capacities of the effective channel $\map C_{\rm eff}$ arising from the use of a bistochastic channel $\map C$ in an indefinite input-output direction.

    \section{Communication advantages of indefinite input-output direction}
    \label{sec:advantages}
    
    Here we apply our communication model to two concrete examples of noisy qubit channels:  depolarising channels and dephasing channels. In both examples, we will observe  advantages of input-output indefiniteness in the transmission of classical and quantum data with respect to the basic scenario where the channels are used in a definite direction.

\begin{figure}
        \centering
        \includegraphics[]{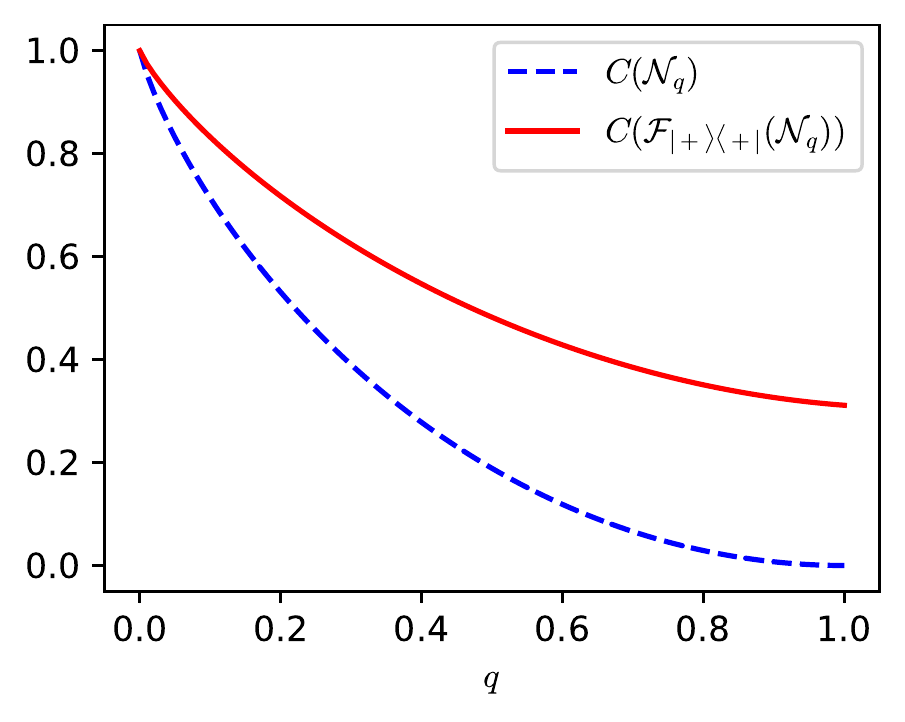}
        \caption{Classical capacity enhancement for  qubit depolarising channels. The lower line (blue, dashed) and top line (red, solid) represent the classical capacity of $\map N_q$ and the effective channel $\map F_{ |+\>\<+|}  (\map N_q)$, respectively, for all possible values of the depolarisation probability $q$.}
        \label{fig:classical_capacity}
    \end{figure}

    \subsection{Enhanced transmission of classical information  through depolarising channels}
    
A qubit depolarising channel  with depolarisation probability $q$ is a linear map  $\map N_q:  M_2  (\C) \to M_2 (\C)$ of the form
    \begin{align}
    \nonumber    \map N_q(\rho)  &= q \,  \frac{I}{2}+   (1-q) \rho   \\
        &=  \left( 1-q  +   \frac q 4  \right) \,\rho  +  \frac q4  \,  (X\rho  X  +  Y\rho  Y  +  Z\rho  Z )  \, ,   \label{eq:qubitdepolarising}
    \end{align}
    where $X,Y,Z$ are the three Pauli matrices. Like all Pauli channels, the depolarising channel  is bistochastic and invariant under transposition. 
     
Now, suppose that Charlie uses a depolarising channel to set up a  communication link between Alice and Bob, as in the model of Fig. \ref{fig:model}. 
 Charlie's optimal strategy, provided by Proposition \ref{prop:optimalprotocol}, results into the effective channel  (\ref{Ceffshort}), which in this case reads  
           \begin{equation}
        \label{eq:qubitflipdepolar}
        \big[\map F_{|+\>\<+|}  ( \map N_q ) \big]  \,  (\rho) =  \left(1   -    \frac q 4\right) \map C_+  (\rho)\otimes |+\>\<+| + \frac q 4  \,  \map C_- (\rho)    \otimes  |-\>\<-|  \, ,
    \end{equation}
    with  \begin{align}
        \map C_+  (\rho)  :  =  \frac {  4  ( 1-q)  \, \rho  +   q (  \rho  +  X \rho  X  +  Z \rho Z)}{4-q}     \qquad {\rm and}  \qquad  \map C_- (\rho)  :=  Y  \rho  Y \, .
    \end{align}

We now  compare the classical capacities of the channels    
$\map N_q$ and $\map F_{ |+\>\<+|} (\map N_q)$.  
The classical capacity of channel  $\map N_q$ is well-known   \cite{king2003capacity}, and is given by 
\begin{align}
C(\map N_q) = 1 + \left( 1-\frac q 2 \right) \log \left( 1-\frac q 2 \right) + \frac q 2 \log \left( \frac q 2 \right)\, ,
\end{align}
where $\log$ denotes the logarithm in base 2.  

The classical capacity  of the channel $\map F_{ |+\>\<+|} (\map N_q)$ is derived in Appendix \ref{app:holevo}, where we also address the case of general target systems of dimension $d\ge 2$.  There, we prove that the classical capacity of the channel $\map F_{|+\>\<+|}  (\map N_q)$ is a convex combination of the capacities of the channels $\map C_+$ and $\map C_-$, and takes the value
    \begin{align}
C\big(\map F_{ |+\>\<+|}  (\map N_q)\big)  
          =  C  (\map N_q)   -  \frac q4  \, \log q - \left(1 - \frac q 4 \right) \log \left(1 - \frac q 4 \right)  \, .         \label{qubitdepcapacity}
    \end{align}

This equation shows that the classical capacity  $C \big (\map F_{|+\>\<+|}  (\map N_q)\big)$ is strictly larger than the classical capacity of the original depolarising channel $\map N_q$  for every positive value of the depolarisation probability. A comparison between the capacities of the two channels is  illustrated in  Fig. \ref{fig:classical_capacity}.  In the most extreme case, $q=1$,  input-output indefiniteness  in the channel $ \map F_{|+\>\<+|}  (  \map N_1)$ permits communication with a capacity of $0.3113$ bits per channel use, even if the original channel $\map N_1$  has zero classical capacity.  
 
 The $0.3113$ capacity achieved with input-output indefiniteness  can be compared with the maximum capacity achievable by other communication protocols using coherent superpositions of configurations. For example, when two completely depolarising channels are used in a superposition of two alternative orders, the maximum classical capacity is 0.049 bits \cite{ebler2018enhanced,chiribella2021quantum}.  If instead two completely depolarising channels are placed on two alternative paths, and a quantum particle is sent through these paths in a quantum superposition,  then  the maximum classical capacity is $0.16$ bits \cite{abbott2018communication,kristjansson2021witnessing}.  
 Remarkably,  comparison between these setups shows that a single channel in a  superposition of two input-output directions can  offer   larger benefits than two channels in a superposition of two orders, or in a superposition of two paths.

    \begin{figure}
        \centering
        \includegraphics[]{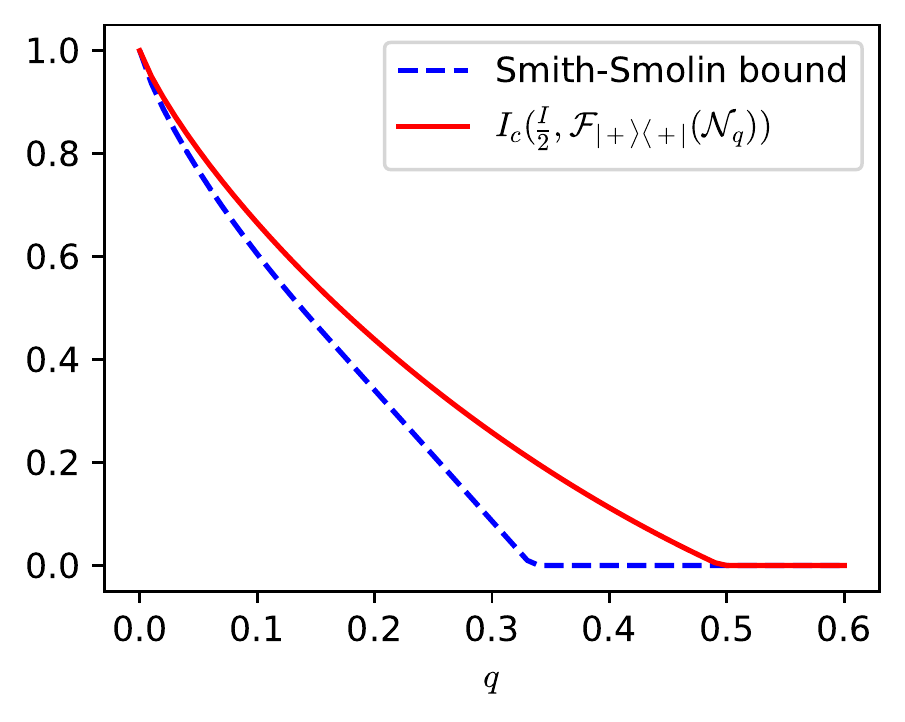}
        \caption{Quantum capacity enhancement for the qubit depolarising channel $\map N_q$. The lower line (blue, dashed) represents the Smith-Smolin upper bound on $Q(\map N_q)$. The top line (red, solid) represents the coherent information of the effective channel $\map F_{ |+\>\<+|}  (\map N_q)$ when acting on the the maximally mixed state $\frac I 2$. This value serves as a lower bound of $Q(\map F_{ |+\>\<+|}  (\map N_q))$.}
        \label{fig:quantum_capacity}
    \end{figure}

    \subsection{Enhanced transmission of quantum information   through depolarising channels}
    We now consider the transmission of quantum data  through depolarising channels in a superposition of input-output directions.  One challenge here is that no close-form expression for the quantum capacity of the depolarising channel has been  found up to date.  
    In order to establish an advantage of input-output indefiniteness, we will  use an upper bound on the quantum capacity, derived by Smith and Smolin in Ref. \cite{smith2008additive}.  There, they showed that the quantum capacity $Q  (\map N_q)$ is upper bounded as     
\begin{align}\label{eq:ss}
  Q(\map N_q)  \le  \op{co}\left[ 1-H(p), H\left( \frac{1-\gamma(p)}{2} \right) - H\left( \frac{\gamma(p)}{2} \right), 1-4p \right] \, ,
\end{align} 
where $p = 3q/4$, $\gamma(p) = 4\sqrt{1-p}(1-\sqrt{1-p})$, $H(p) = -p\log p - (1-p)\log (1-p)$ and $\op{co}[f_1(p), \cdots, f_n(p)]$ denotes the maximal convex function that is less than or equal to all $f_i(p)$, $i=1, \cdots, n$.

We  now compare this upper bound with a lower bound on  the quantum capacity of the effective channel $\map F_{ |+\>\<+|}  (\map N_q)$.      By Eq. (\ref{boundQ}),  the quantum capacity $Q (\map F_{ |+\>\<+|}  (\map N_q))$ is lower bounded by  the  coherent information $I_{\rm c}  (\map F_{ |+\>\<+|}  (\map N_q)) $, which in turn can be lower bounded  by setting $\rho  =  I/2$  in Eq.  (\ref{cohinfo}).  The resulting lower bound is 
       \begin{align}\label{qcapacitydep}
  Q  \big(  \map F_{|+\>\<+|}   (\map N_q)\big) \ge   - \left(1- \frac q {4}  \right) \log \left(\frac 1 2 - \frac q {8} \right) -  \frac q {4} \, \log \left( \frac q {8} \right) +   \frac{3q}{4} \,  \log \left(\frac{q}{4}\right) + \left(1-\frac{3q}{4}\right) \log \left(1-\frac{3q}{4}\right)  \, .
    \end{align}
The derivation of this equation is provided in  Appendix \ref{app:coherentinfo}, where we also extend the bound to general quantum systems of dimension $d\ge 2$.

Comparing Eqs.  (\ref{eq:ss}) and   (\ref{qcapacitydep}), we obtain that the strict inequality     $  Q  \big(  \map F_{|+\>\<+|}   (\map N_q)\big)  >  Q (\map N_q)$  is valid for every value of $q$ in the interval $(0, 0.495)$.  Our lower bound on the quantum capacity is plotted in Fig. \ref{fig:quantum_capacity}, along with the Smith-Smolin upper bound on the quantum capacity of the depolarising channel. 
  
When the depolarising probability reaches the value $q  =  1/3$,  the Smith-Smolin bound implies that the quantum capacity of the depolarising channel  $\map N_q$ is zero. In contrast,  our lower bound shows that  the quantum capacity of the  flipped depolarising channel $\map F_{|+\>\<+|}  (\map N_q)$  is  at least 0.2063 qubits at $q=1/3$, and remains larger than zero until $q$ is approximately $0.5$.    

Note that a non-zero quantum capacity also implies a non-zero private capacity $P\big(\map F_{|+\>\<+|}  (\map N_q)\big)$, due to the bound (\ref{privatebound}). Hence, the flipped depolarising channel $\map F_{|+\>\<+|}  (\map N_q)$  can be used to securely transmit private messages from Alice to Bob.  Notably, the definition of private capacity guarantees that the transmission is secure even if the eavesdropper  has access to the whole environment of the channel $\map F_{|+\>\<+|}  (\map N_q)$, which includes in particular the case in which the eavesdropper has access to Charlie's measurement outcome.  

In summary, input-output indefiniteness offers the possibility to achieve a reliable transmission of quantum states  starting from depolarising channels that have zero quantum capacity. Remarkably, this feature cannot be achieved by using two depolarising channels in a superposition of alternative orders: as shown in Ref. \cite{chiribella2021quantum}, the quantum capacity obtained by placing two depolarising channels in a superposition of orders is non-zero only if the quantum capacity of the original channels is non-zero.   In contrast, the superposition of two alternative input-output directions permits quantum communication even with a single depolarising channel with zero quantum capacity. 

The increase of quantum capacity from zero to nonzero enables the use of  quantum key distribution (QKD) protocols, such as BB84  \cite{Bennett85} and E91  \cite{Ekert91},  or the use of quantum secure direct communication (QSDC) protocols  \cite{long2002theoretically,deng2003two,long2007quantum} to transmit private messages from the sender to the receiver (see also \cite{zhang2022realization} for a recent experimental demonstration and \cite{long2022evolutionary,sheng2022one,zhou2023device} for recent theoretical developments  on this topic).  In principle, these protocols can be implemented by encoding the appropriate quantum states in the multipartite states that achieve the quantum capacity $ Q  \big(  \map F_{|+\>\<+|}   (\map N_q)\big)$.   In practice, however, this approach is still challenging because the capacity-achieving encoding is not known in general, and may involve large multipartite states that are hard to implement. Fortunately, a simpler approach is also possible: indeed, the superposition of input-output directions also enables a {\em noiseless heralded transmission of quantum data}. As one can see from Eq. (\ref{eq:qubitflipdepolar}), the evolution of the target system is unitary whenever the control system is found in the state $|-\>$. This means that, when Bob receives the outcome -, he can retrieve the state of  Alice's qubit without any error. Such heralded transmission allows Alice and Bob to implement QKD and QSDC protocols  using  the quantum states of the original protocols, which are much simpler to realise compared to  their encoded version.

     
    \subsection{Perfect transmission of quantum information through dephasing channels}
    
   We now show a scenario where input-output indefiniteness enables a perfect, deterministic transmission of quantum information.  Consider a qubit dephasing channel of the form
     \begin{equation}
        \map M_p= (1-p) \rho + p Y\rho Y \, .
    \end{equation}
    In normal conditions, this channel $\map M_p$ can be used to transmit a  classical bit,  by encoding  it into an  eigenstate of the Pauli-$Y$ gate. However,  transmitting a quantum bit per channel use of  $\map M_p$ is  impossible except in the extreme cases  $p=0$ or $p=1$, in which the channel $\map M_p$ is noiseless.  Indeed,  the quantum capacity of the dephasing channel $\map M_q$ is \cite{wilde2013quantum}         \begin{equation}
        Q(\map M_p) = Q^{(1)}(\map M_p) = 1 - H(p) \, ,
    \end{equation}
    and is strictly smaller than 1 for every  $q\in  (0,1)$. 
    In stark contrast, we now show that using the channel $\map M_q$ in a coherent superposition of the forward and backward direction yields a quantum capacity of 1 qubit per channel use, for every possible value of $q$.  
    
Suppose that the depolarising channel $\map M_q$ is used in the scenario of Fig. \ref{fig:model},  with the control qubit initialised in  the maximally coherent state $\omega$. The resulting channel $\map F_{|+\>\<+|}(\map M_p)$ can be easily computed from Eq. (\ref{simplified+}), which yields
         \begin{equation}
       \big [ \map F_{|+\>\<+|}  (  \map M_p)\big]  \,  (\rho) = (1-p) \rho \otimes |+\>\<+| + p Y\rho Y \otimes |-\>\<-| \, .  
    \end{equation}
    Using this channel, Alice can transmit a qubit to Bob deterministically and without error. To retrieve Alice's input state $\rho$,    Bob needs only to  measure  the control system in the Fourier basis $\{|+\>,  |-\>\}$: if the measurement outcome is ``$+$'', then Bob obtained $\rho$ directly; if the outcome is ``$-$'', then Bob can apply Pauli-$Y$ gate to the state to recover $\rho$. 
    
    Summarising, the flipped channel $\map F_{|+\>\<+|}  (\map M_p)$ offers an advantage over the original dephasing channel $\map M_p$ for every  value of the dephasing probability except $p=0$ and $p=1$.  The most extreme advantage arises for $p=1/2$,      when the original dephasing channel  has zero quantum capacity.

The possibility of perfect quantum communication through zero-capacity  channels was also observed for quantum communication with indefinite causal order  \cite{Chiribella2021Indefinite}, where it was later shown to enable a task called  random-receiver quantum communication \cite{bhattacharya2021random}.  The same task can be achieved also in the present scenario, using a single dephasing channel in a coherent superposition of two alternative input-output directions instead of two channels in a coherent superposition of two alternative orders.

    \section{Conclusions}
    \label{sec:summary}

    In this paper, we developed a  communication model where bidirectional quantum devices can be used in an indefinite input-output direction, and we showed that this input-output indefiniteness  can enhance the classical and quantum capacity of noisy channels.   These enhancements are similar to those arising from indefiniteness of the path of quantum particles through noisy processes, or from indefinite of the order in which the noisy processes occur.  On the other hand, input-output indefiniteness  does not require the use of multiple  channels, or the use of a given  channel at multiple moments of time.  Moreover, the results of this paper indicate that  indefiniteness of the input-output direction  generally offers stronger advantages compared to the above models. For example, a single completely depolarising channel used in a superposition of two directions can reach a classical capacity of 0.31 bits, while two completely depolarising channels  can at most reach a capacity of $0.16$ bits when put on two alternative paths, and of $0.049$ bits when used in two alternative  orders.

The study of quantum communication with indefinite input-output direction sheds light on the operational features of the recently introduced ``quantum time flip'' operation.  In turn, the quantum time flip and other operations with indefinite input-output direction can be used as a mathematical framework to formulate  hypothetical scenarios where the arrow of time is in a quantum superposition. While these scenarios are far from the currently known physics, and therefore necessarily speculative, they can help clarifying the consequences of basic quantum principles, when applied to unusual settings involving the order of events in time.   For example, consider a thought experiment where  two parties, Alice and Bob, operate inside two closed laboratories  and interact with each other only through a quantum channel, used by Alice to send quantum states to Bob.   If the channel is bidirectional and the laws of nature are time-symmetric, this scenario  is in principle  compatible with two different time orders between Alice's and Bob's operations: Alice and Bob could operate in the forward order (with Alice preparing input states   in Bob's past, and the signals propagating from Alice's laboratory to Bob's laboratory in the forward time direction) or in the backward order  (with Alice preparing  input states in Bob's future, and the signals propagating from Alice's laboratory to Bob's laboratory in the backward time direction).  More generally, the time  order of Alice's and Bob's operations may be indefinite, as in the quantum time flip.    This closed laboratory situation is analogue to the situation considered by Oreshkov, Costa, and Brukner  in the study of indefinite causal order \cite{Oreshkov12}, with the only difference that here the causal order between Alice and Bob is well defined (Alice can send signals to Bob, but not {\em vice-versa}) while the time order is indefinite.    While the physical realisation of these scenarios is currently an open problem, the mathematical framework of quantum operations with indefinite input-output direction provides a rigorous way to explore  them  at a conceptual level.  In addition, the superposition of  time directions can be simulated in table top experiments by coherently controlling the path of single photons, as envisaged in \cite{chiribella2022quantum} and recently demonstrated in two experiments \cite{guo2022experimental,stromberg2022experimental}.

Another direction of future research is the experimental demonstration of the communication scenarios explored in this paper.  
Notably, all our communication protocols    can already be demonstrated with a simple adaptation of the photonic setups  in Refs. \cite{guo2022experimental,stromberg2022experimental}. Indeed, these setups reproduced the action of the quantum time flip on  unitary qubit channels, by controlling the direction in which a single photon traverses an arbitrary polarisation rotator. By randomising the choice of unitary channel, one can then realise the flipped Pauli channels appearing in our paper.       The experimental realisation of quantum communication with indefinite input-output direction  is expected to contribute to the development of a toolbox for quantum control over the configuration of multiple quantum devices, which may prove technologically useful in a longer term. 

In the shorter term, our work provides the starting point for a number of foundational explorations. First,  an interesting direction is the investigation of  scenarios where  both the input-output direction and the causal order of multiple processes are  subject to quantum indefiniteness. A mathematical framework for these more general operations was recently established in \cite{chiribella2022quantum}, but very little is currently known about their information-theoretic potential. The angle of quantum communication, explored in this paper, represents a promising approach to explore the capabilities of new operations with indefinite order and direction. Another interesting area of  future research  is the study of quantum thermodynamic tasks assisted by indefinite input-output direction. Recently, the communication advantages of indefinite causal order stimulated new research  in quantum thermodynamics \cite{felce2020quantum,Guha20,simonov2022work}.  Similarly,  the communication advantages  provided in this paper suggest  new thermodynamic protocols where the input-output direction of one or more processes is indefinite.

    \section{Acknowledgements}
    The authors thank H.Kristjánsson for helpful discussions.
This work is supported by the Hong Kong Research Grant Council through grant 17307719  and though the Senior Research Fellowship Scheme SRFS2021-7S02, by the Croucher Foundation,  
and by the ID No. 62312 grant from the John Templeton Foundation, as part of the ‘The Quantum Information Structure of Spacetime, Second phase (QISS 2)’ Project (The Quantum Information Structure of Spacetime (QISS), Second Phase - John Templeton Foundation).  Research at the Perimeter Institute is supported by the Government of Canada through the Department of Innovation, Science and Economic Development Canada and by the Province of Ontario through the Ministry of Research, Innovation and Science. The opinions expressed in this publication are those of the authors and do not necessarily reflect the views of the John Templeton Foundation.

    \bibliography{references}

    \appendix

\section{Proof that the communication model in Fig. \ref{fig:model} does not allow for communication independently of the input channels}
\label{app:proof}

Here we show that our communication model does  not contain side-channel generating operations, defined as follows: 
    \begin{defi} [Side-channel generating operations]
    \label{def:side}
Let  $\set M_{\rm free} \subset \set M$ be the set of free supermaps in a given resource theory of communication,  $\set M$ being  the set of all quantum supermaps. A supermap $\map S \in \set M$  {\em generates a classical (quantum) side-channel} if there exist two free supermaps $\map S_{1}  \in \set{M}_{\rm free}$ and $\map S_{2} \in \set{M}_{\rm free}$ such that
        \begin{equation}
        \label{eq:nsc}
        (\map S_{2}  \circ \map S  \circ \map S_{1} )( \map C )  = \map C_0  \, ,
        \end{equation}
        where  $\map C$ is an arbitrary input channel and $\map C_0$ is a fixed quantum channel with non-zero classical (quantum) capacity.   
    \end{defi}

In the following,  we  show that no  supermap of the form (\ref{eq:generalfree}) can generate a side-channel in the sense of Definition \ref{def:side}.   Specifically, we will prove the following theorem:
  \begin{theo}\label{theo:noside}
  For  every supermap $\map S$ of the form  (\ref{eq:generalfree}), if the supermap $\map S$ satisfies the condition $\map S (  \map C)=\map C_0$ for a fixed $\map C_0$ and for every bistochastic channel $\map C$,  then $\map C_0$ has zero classical (quantum) capacity.
  \end{theo}

 To prove the  theorem, we need to specify the set of free supermaps in our model.     Consider first  the supermap $\map F_\omega$ that  transforms an initial bistochastic channel $\map C$ into a new channel $\map F_\omega  (\map C)$, defined as 
    \begin{equation}
        \label{eq:effective}
        \begin{split}
       \left[ \map F_{\omega}  (\map C)\right](\rho)
        &= \map F(\map C)(\rho \otimes \omega) \\
        &= \sum_{i=1}^r (C_i \rho C_i^\dagger) \otimes  |0\>\<0|\omega  |0\>\<0|  
        + \sum_{i=1}^r (C_i \rho \overline C_i) \otimes|0\>\<0|   \omega |1\>\<1| \\
        &\quad + \sum_{i=1}^r (C_i^T \rho C_i^\dagger) \otimes  |1\>\<1|   \omega |0\>\<0|
        + \sum_{i=1}^r (C_i^T \rho \overline C_i) \otimes |1\>\<1| \omega |1\>\<1| \, ,
        \end{split}
    \end{equation}
    where  $\rho$ is the quantum state of the target system, and  $\{\map C_i\}$  are the  Kraus operator of  the bistochastic channel  $\map C$. 
  We call the supermap $\map F_\omega$ the {\em time-flip placement} with control state $\omega$.  Physically, this supermap replaces the original channel $\map C$ with a new channel $\map F_\omega (\map C)$, in which the the forward and backward processes $\map C$ and $\map C^T$  are combined in a way determined by the state $\omega$ of the control system.

 Since in our model the communication provider measures the control system and communicates the outcome to  Bob,  the basic type of  free supermap is of the form  
 \begin{align}\label{eq:QTFMAP}
\map S_{\omega,  \map M}   (\map C)     =   (\map I_{\rm target} \otimes \map M)   \circ \map F_{\omega}  (\map C) \, ,      
 \end{align}
 where $\map I_{{\rm target}}$ is the identity channel acting on the  target system and $\map M$ is a quantum-to-classical channel \cite{wilde2013quantum}  transforming  the  control system into a classical register sent to Bob.   Explicitly, $\map M$ must have the  form 
  $\map M (\omega)   =  \sum_{i=1}^k   \Tr  [  P_i \,  \omega]  \,  |i\>\<i|$, where $(P_i)_{i=1}^k$ are operators describing Charlie's measurement  ($P_i\ge 0$ for every $i$ and $\sum_{i=1}^k   P_i   =  I_{{\rm control}}$,  $I_{{\rm control}}$ being the identity on the  control system), and $(|i\>)_{i=1}^k$ are orthogonal states of a classical register.

  The set of free supermaps in our communication model consists of all possible  time-flip placements, combined with all possible measurements on the control system, and all  local encoding and decoding operations performed by Alice and Bob.  In formula, the most general free supermap considered in our model is of the form 
  \begin{align}\label{eq:generalfree}
      \map S (  \map C)    =   \map D \circ  \map S_{\omega, \map M} (\map C )   \circ \map E \, ,   
  \end{align}
  where $\map E$ ($\map D$) is a quantum channel (completely positive trace-preserving map) used by Alice (Bob) to  encode (decode) information.

The composite channel $\map S (  \map C)$ as defined in Eq. \ref{eq:generalfree} can be re-expressed as (using Eq. \ref{eq:QTFMAP}) $\map S (  \map C)=  \map D' \circ \map F_{\omega}  (\map C) \circ \map E $, where $\map D' =\map D \circ (\map I_{\rm target} \otimes \map M)$. This is more general than the original setting of Charlie measuring the control bit and sending the result to Bob classically. To show that no side-channels generated, We require the output channel independent of the choice of the bistochastic channel $\map C$ and has zero capacity. The key point of the proof is that the decoding channel $\map D'$ is a linear map.

In the following, we will denote by $\set{SymChan}$ the set of channels which have a Kraus representation consisting of symmetric matrices. The quantum time flip has no effect on the channels from $\set{SymChan}$ because $\map F_\omega(\map C) = \map C \otimes \omega$ for every channel $\map C$ from $\set{SymChan}$.

\begin{lemma}
    \label{theo:independent}
    Let $\map E \in \Chan(A, B)$ be a quantum channel, and let $\rho$ be a quantum state on $\spc H_A$. If $\map E(\map C(\rho))$ is independent of the choice of $\map C \in \set{SymChan}(A)$, then $\map E(\map C(\rho)) = \map E(I/d)$, where $d$ is the dimension of $\spc H_A$.
\end{lemma}

\Proof
Firstly we consider a class of projection channels $\map C_{\theta, m, n}$ with a real parameter $\theta$ and integers $m, n \in \{ 0, \cdots, d-1 \},\ (m \neq n)$. The Pauli matrices on the subspace $\Span\{ |m\>, |n\> \}$ are defined as follows:
\begin{equation}
\begin{split}
    &I_{m,n} = |m\>\<m| + |n\>\<n| \, , \quad X_{m,n} = |m\>\<n| + |n\>\<m| \, , \\
    &Y_{m,n} = i|m\>\<n| - i|n\>\<m| \, , \quad Z_{m,n} = |m\>\<m| - |n\>\<n| \, .
\end{split}
\end{equation}
And the channel $\map C_{\theta, m, n}$ is given by:
\begin{equation}\label{eq:channeldef}
    \map C_{\theta, m, n}(\cdot) = P_{\theta,m,n,0} (\cdot) P_{\theta,m,n,0} + P_{\theta,m,n,1} (\cdot) P_{\theta,m,n,1} + (I-I_{m,n})(\cdot)(I-I_{m,n}) \, ,
\end{equation}
with $P_{\theta,m,n,0}$, $P_{\theta,m,n,1}$ being the following projectors on the subspace $\Span\{|m\>, |n\>\}$
\begin{equation}
\label{eq:projectors}
\begin{split}
    P_{\theta,m,n,0} = \frac 1 2 [ (1+\sin\theta)|m\>\<m| + \cos\theta|m\>\<n| + \cos\theta |n\>\<m| +( 1-\sin\theta )|n\>\<n| ] \, , \\
    P_{\theta,m,n,1} = \frac 1 2 [ (1-\sin\theta)|m\>\<m| - \cos\theta|m\>\<n| - \cos\theta |n\>\<m| +( 1+\sin\theta )|n\>\<n| ] \, . 
\end{split}
\end{equation}

The output of the channel $\map E (\map C_{\theta, m, n})$ is
\begin{equation}
    \rho' = \map E(\map C_{\theta, m, n}(\rho)) \, .
\end{equation}
The state $\rho$, when projected to the subspace $\Span\{|m\>, |n\>\}$, can be decomposed as follows:
\begin{equation}
    \frac{\Tr(I_{m,n}\rho)}{2}(I_{m,n} + r_1X_{m,n} + r_2Y_{m,n} + r_3Z_{m,n}) \, .
\end{equation}
where $r_1, r_2, r_3$ are real numbers such that $r_1^2 + r_2^2 + r_3^2 \leq 1$. And the projections by $P_{\theta,m,n,0}$ and $P_{\theta,m,n,1}$ is given by
\begin{multline}\label{eq:projection}
    P_{\theta,m,n,0} \rho P_{\theta,m,n,0} + P_{\theta,m,n,1} \rho P_{\theta,m,n,1} = \frac {\Tr(I_{m,n}\rho)} 2 \Big( I_{m,n} + \frac 1 2 r_1X_{m,n} + \frac 1 2 r_3Z_{m,n} \\ + \frac 1 2 (r_1\cos(2\theta) + r_3\sin(2\theta)) X_{m,n} + \frac 1 2 (r_1\sin(2\theta) - r_3\cos(2\theta)) Z_{m,n} \Big) \, . 
\end{multline}
The condition that $\rho'$ is independent of the channel parameters $\theta, m, n$ implies that
\begin{equation}
\begin{split}\label{eq:differentiation}
    \frac{\partial \rho'}{\partial \theta} 
    &= \frac{\partial \map E(P_{\theta,m,n,0} \rho P_{\theta,m,n,0} + P_{\theta,m,n,1} \rho P_{\theta,m,n,1})}{\partial \theta} \\
    &= \map E \big( \frac {\Tr(I_{m,n}\rho)} 2 ((-r_1\sin(2\theta) + r_3\cos(2\theta)) X_{m,n} + (r_1\cos(2\theta) + r_3\sin(2\theta)) Z_{m,n}) \big) \\
    &= 0 \, . \quad \forall \theta, m, n \, (m \neq n)
\end{split}
\end{equation}
It follows that $\forall m, n \,  (m \neq n)$
\begin{equation}
    \label{eq:null}
     \Tr(I_{m,n}\rho) \cdot \sqrt{r_1^2 + r_3^2} \cdot \map E (X_{m,n}) = 0 \, , \quad \Tr(I_{m,n}\rho) \cdot \sqrt{r_1^2 + r_3^2} \cdot \map E (Z_{m,n}) = 0 \, .
\end{equation}
Eq. \ref{eq:projection} and \ref{eq:null} imply that
\begin{equation}
    \map E(P_{\theta,m,n,0} \rho P_{\theta,m,n,0} + P_{\theta,m,n,1} \rho P_{\theta,m,n,1}) = \Tr(I_{m,n}\rho)\map E\left(\frac {I_{m,n}} 2 \right) \, . \quad \forall \theta, m, n (m \neq n)
\end{equation}
Combined with the following equation,
\begin{equation}
    \map E(P_{\pi/2,m,n,0} \rho P_{\pi/2,m,n,0} + P_{\pi/2,m,n,1} \rho P_{\pi/2,m,n,1}) = \<m|\rho|m\>\map E(|m\>\<m|) + \<n|\rho|n\>\map E(|n\>\<n|) \, ,
\end{equation}
we can see that for arbitrary $m,n$ from $\{ 0, \cdots, d-1 \}$, it holds that
\begin{equation}
    \<m|\rho|m\>\map E(|m\>\<m|) + \<n|\rho|n\> \map E(|n\>\<n|)) = \<m|\rho|m\>\map E(|n\>\<n|) + \<n|\rho|n\>\map E(|m\>\<m|) \, .
\end{equation}

Secondly we consider another channel $\map C_0$ defined as
\begin{equation}
    \map C_0(\cdot) = \frac 1 {2d} \sum_{m=0}^{d-1}\sum_{n=0}^{d-1} |m\>\<m|(\cdot)|m\>\<m| + |n\>\<n|(\cdot)|n\>\<n|.
\end{equation}
Noticing that the Kraus operators of $\map C_0$ are also symmetric matrices, we have
\begin{align}
    \rho' 
    &= \map E(\map C_0(\rho)) \nonumber \\
    &= \frac 1 {2d} \sum_{m=0}^{d-1}\sum_{n=0}^{d-1} \<m|\rho|m\>\map E(|n\>\<n|) + \<n|\rho|n\>\map E(|m\>\<m|) \nonumber \\
    &= \map E\left( \frac I d \right) \, . \label{eq:fixedoutput}
\end{align}
\qed

Now we prove Theorem \ref{theo:noside} with Lemma \ref{theo:independent}.

\medskip  
{\bf Proof of Theorem \ref{theo:noside}.}
Suppose that there exists $\map D'$ and $\map E$ such that $\map D' \circ \map F_\omega(\map C) \circ \map E$ is a fixed channel for every bistochastic channel $\map C$. Considering only the channels from $\set{SymChan}$, Lemma \ref{theo:independent} implies that
\begin{equation}
    \forall \rho \, \forall \map C \in \set{SymChan} \, , \quad (\map D' \circ \map F_\omega(\map C) \circ \map E)(\rho) = \map D' (\map C(\map E(\rho)) \otimes \omega) = \map D' \left(\frac I d \otimes \omega \right) \, .
\end{equation}
In conclusion, the composite channel $\map D' \circ \map F_\omega(\map C) \circ \map E$, whose output is a fixed state, has no capacity. Thus the communication model does not generate side-channels. \qed

\medskip  

  In summary, we have shown that our communication model does not allow Alice and Bob to communicate independently of the channel $\map C$ used by Charlie to set up the communication link between them.  Hence, our communication model satisfies the minimal requirement put forward in Ref. \cite{kristjansson2020resource}.    It is worth stressing, however, that this  requirement {\em per se}  is not   a sufficient condition for a resource theory of communication to be meaningful, but rather serves as a sanity check to  rule out resource theories  exhibiting trivial advantages. Ultimately, the motivations to study a communication model should come from other considerations. In our case, the motivation of the communication model is to study the information-theoretic capabilities of the quantum time flip, and, more generally, to gain an insights in the potential of indefinite input-output direction in quantum information tasks.

\section{Characterisation of the  transposition invariant channels}\label{app:invariant}  

Here we provide the proof of Proposition \ref{prop:invariant}:  a quantum channel $\map C$ admits a Kraus decomposition with only symmetric or anti-symmetric operators if and only if it is invariant under transposition.  

One direction is straightforward: if every Kraus operator $C_i$ satisfies either $C_i^T  = C_i  $ or $C_i^T  = -  C_i$, then the transpose channel   $\map C^T :  \rho  \mapsto  \sum_i  C_i^T  \rho  \overline C_i$ coincides with the original channel $\map C$.  

To prove the converse direction, we use the Choi representation \cite{choi1975completely}, which provides a one-to-one correspondence between linear maps and bipartite operators. For the quantum channel $\map C$, the Choi operator ${\sf Choi} (\map C)$ is given by 
\begin{align}
{\sf Choi}  (  \map C)   :=   (\map C  \otimes \map I)   (  |I\kk\bb I)   =   \sum_i  \, |C_i\kk \bb C_i|  \, ,    
\end{align}
where $|I\kk :  =  \sum_n  |n\> \otimes |n\>$  is the canonical (unnormalised) maximally entangled state, and 
\begin{align}\label{doubleket}
    |C_i\kk  :  =  (C_i\otimes I)  |I\kk \, .
\end{align}  In the Choi representation,  the transposition of the channel corresponds to a swap of the input and the output systems \cite{chiribella2021symmetries,chiribella2022quantum}:   
\begin{align}
 {\sf Choi}  (  \map C^T)   :=   (\map C^T  \otimes \map I)   (  |I\kk\bb I)   =  {\tt SWAP}~ {\sf Choi}  (  \map C) ~  {\tt SWAP} \, ,        
\end{align}
where $\tt SWAP$ is the swap operator, defined by the relation ${\tt SWAP}\,  \left(  |\phi\>\otimes |\psi\>\right)  =  |\psi\> \otimes |\phi\>$ for every pair of vectors $|\phi\>$ and $|\psi\>$.  

Hence, the quantum channel $\map C$ is invariant under transposition if and only if  
\begin{align}
    {\tt SWAP}~ {\sf Choi}  (  \map C) ~  {\tt SWAP}  =  {\sf Choi}  (  \map C)\, ,
\end{align}
that is, if and only if $ {\sf Choi}  (  \map C)$ commutes with the swap operator ${\tt SWAP}=0$.   The condition implies that the Choi operator $ {\sf Choi}  (  \map C)$ has a spectral decomposition   in which every eigenvector  belongs to an eigenspace of the swap operator.  Using the double ket notation (\ref{doubleket}), such a spectral decomposition can be written as $ {\sf Choi}  (  \map C)  =  \sum_j  |\Psi_j\kk\bb\Psi_j|$, with $|\Psi_j\kk  =  (\Psi_j \otimes I)  \,  |I\kk$ for some suitable operator $\Psi_j$.   Since the eigenvectors of the swap operator are $+1$ and $-1$,  each  eigenvector $|\Psi_j\kk$ must satisfy either the condition ${\tt SWAP}  \,  |\Psi_j\kk  = |\Psi_j\kk$   or the condition ${\tt SWAP}  \,  |\Psi_j\kk  = -|\Psi_j\kk$.    On the other hand, a basic property of the double ket notation is that  $ {\tt SWAP}  \,  |\Psi_j  \kk  =    |\Psi_j^T \kk$ \cite{royer1991wigner,dariano2000bell}.   Hence, the operator $\Psi_j$ must satisfy either the condition $\Psi_j^T =  \Psi_j $ or the condition $\Psi_j^T  =  - \Psi_j$.   Now, the operators $\{\Psi_j\}_j$ are a Kraus representation of the channel $\map C$, as one can deduce from the relation 
\begin{align}
    (\map C  \otimes \map I)   (  |I\kk\bb I|)     =  {\sf Choi}  (  \map C) =  \sum_j   |\Psi_j \kk \bb \Psi_j  |  =  \sum_j   (\Psi_j \otimes I)  \,  |I\kk \bb   I|  \,   (\Psi_j \otimes I)^\dag  \, ,      
\end{align}
and from the fact that the map ${\sf Choi}$ is one-to-one, which  implies equality of the  map $\map C$ with the map $  \map C' :  \rho  \mapsto  \sum_j \Psi_j   \,  \rho  \,  \Psi_j^\dag $. 
Summarising, the channel $\map C$ has a Kraus representation  $(\Psi_j)_j$ in which every operator is either symmetric or anti-symmetric. This concludes the proof of Proposition \ref{prop:invariant}. 

\section{Derivation of the Holevo information of the effective channel based on the depolarising channel}
\label{app:holevo}

To derive the Holevo information of the effective channel, we first study the properties of labeled random channels defined below. The properties can be applied to the effective channel based on the depolarising channel because it belongs to the class of labeled random channels.

\subsection{Labeled random channels}
\begin{defi} [Labeled random channel]
Let $\{ \map C_l \}$ be a collection of quantum channels. A labeled random channel is defined as a random mixture of the channels $\{ \map C_l \}$ attached with classical labels. Precisely, let $\{p_L(l)\}$ be a probability distribution, a labeled random channel over $\{C_l\}$ with probability distribution $\{p_L(l)\}$ is
\begin{equation}
    \sum_l p_L(l) \map C_l(\cdot) \otimes |l\>\<l|_L \, .
\end{equation}
\end{defi}

\begin{theo} [Holevo information of a labeled random channel]
\label{theo:labeledrandom}
Provided that the Holevo information of the every channel from $\{\map C_l\}$ can be achieved with a fixed ensemble, the Holevo information of a labeled random channel over $\{\map C_l\}$ with probability distribution $\{p_L(l)\}$ is equal to $\sum_l p_L(l) \chi(\map C_l)$. In addition, if Holevo additivity holds among the channels $\{ \map C_l \}$, then the classical capacity of the labeled random channel reduces to its Holevo information.
\end{theo}
\Proof
Suppose that Alice prepare an ensemble $\{ p_M(m), \rho_m \}$ and keeps her choice of state in a classical register $M$, then after sending a state of the ensemble to Bob via the labeled random channel, they share the following joint state:
\begin{equation}
    \xi_{MBL} = \sum_m p_M(m) |m\>\<m|_M \otimes \sum_l p_L(l) \map C_l(\rho_m) \otimes |l\>\<l|_L \, .
\end{equation}
The Holevo information of the joint state is
\begin{equation}
\label{eq:mixed_holevo}
\begin{split}
    \chi(\xi_{MBL}) 
    &= I(M;BL)_\xi \\
    &= H(BL) - H(BL\mid M) \\
    &\quad \text{(Joint entropy theorem \cite{nielsen2010quantum})} \\
    &= \sum_l p_L(l) H \left[\map C_l\left( \sum_m p_M(m) \rho_m \right)\right] + H(p_L) - \sum_m p_M(m) \left( \sum_l p_L(l) H(\map C_l(\rho_m)) + H(p_L) \right) \\
    &= \sum_l p_L(l) [H(B \mid L = l) - H(B \mid M, L = l)] \\
    &= \sum_l p_L(l) I(M;B \mid L=l) \\
    &= \sum_l p_L(l) \chi(\xi_{MB \mid L=l}) \\
    &\leq \sum_l p_L(l) \chi(C_l) \, .
\end{split}
\end{equation}
If the ensemble $\{ p_M(m), \rho_m \}$ maximises $\chi(\xi_{MB \mid L=l})$ for every choice of $l$, then the inequality in \ref{eq:mixed_holevo} can be saturated, which means that the Holevo information of the labeled random channel is $\sum_l p_L(l) \chi(C_l)$.

Now additionally suppose that Holevo additivity holds among the channels $\{ \map C_l \}$. Consider the following $n$ labeled random channels:
\begin{equation}
    \map D_k(\cdot) = \sum_{l_k} p_{L_k}(l_k) \map C_{l_k}(\cdot) \otimes |l_k\>\<l_k|_{L_k} \quad k = 1, \cdots, n \, .
\end{equation}
The tensor product of the $n$ labeled random channels is also a labeled random channel,
\begin{equation}
    \bigotimes_{k=1}^n \map D_k = \sum_{l_1, \cdots, l_n} p_{L_1}(l_1)\cdots p_{L_n}(l_n)  (\map C_{l_1} \otimes \cdots \otimes \map C_{l_n}) \otimes |l_1, \cdots, l_n\>\<l_1, \cdots, l_n| \, .
\end{equation}
The Holevo information of the channel $\map C_{l_1} \otimes \cdots \otimes \map C_{l_n}$ is additive and thus can be achieved by the $n$-fold product of a fixed ensemble. From the previous point, we can see that the Holevo information of the product $\bigotimes_{k=1}^n \map D_k$ is
\begin{equation}
\begin{split}
    \chi\left( \bigotimes_{k=1}^n \map D_k \right) 
    &= \sum_{l_1, \cdots, l_n} p_{L_1}(l_1)\cdots p_{L_n}(l_n)  \chi(\map C_{l_1} \otimes \cdots \otimes \map C_{l_n}) \\
    &= \sum_{l_1, \cdots, l_n} p_{L_1}(l_1)\cdots p_{L_n}(l_n) (\chi(\map C_{l_1}) +  \cdots + \chi(\map C_{l_n})) \, .
\end{split}
\end{equation}

If the $n$ channels are $n$ independent uses of an identical labeled random channel, i.e., the probability distributions $p_{L_k}$ are all identical to a probability distribution $p_L$. Then the classical capacity of the labeled random channel is a direct result of the weak law of large numbers:
\begin{equation}
    \lim_{n \to \infty} \frac{\chi\left( \bigotimes_{k=1}^n \map D_k \right)}{n} = \sum_l p_L(l)\chi(C_l) \, .
\end{equation}
\qed

\subsection{The effective channel based on the depolarising channel}
Now we can compute the Holevo information of the effective channel when the original channel is the depolarising channel $\map N_q$. A Kraus representation of $\map N_q$ is
\begin{equation}
    \left\{ \sqrt{\frac{q}{2d}} |j\>\<i| \right\}_{i, j \in \{0, \cdots, d-1\} } \cup \{\sqrt{1-q}I\} \, .
\end{equation}

For an input state $\rho$, the effective channel gives the output
\begin{align}
    \map F_{ |+\>\<+|}  (\map N_q) (\rho)
    &= \map F(\map N_q)(\rho \otimes |+\>\<+|) \\
    &= \frac q {2d} \left[ \sum_{ij} (|j\>\<i| \otimes |0\> + |i\>\<j| \otimes |1\>) \rho (|i\>\<j| \otimes \<0| + |j\>\<i| \otimes \<1|) \right] + (1-q)\rho \otimes |+\>\<+|_C \\
    &= \frac q {2d} (I \otimes I + \rho^T \otimes (|0\>\<1|+|1\>\<0|)) + (1-q)\rho \otimes |+\>\<+| \\
    &= \left(\frac{qI}{2d} + \frac q {2d} \cdot \rho^T + (1-q)\rho\right) \otimes |+\>\<+| + \left(\frac{qI}{2d} - \frac q {2d} \cdot \rho^T\right) \otimes |-\>\<-| \, ,
\end{align} 
We can see that the effective channel $\map F_{ |+\>\<+|}  (\map N_q)$ is a labeled random channel over $\{ \map N_q^+, \map N_q^- \}$ with probability distribution $\{ 1-q/2+q/2d, \ q/2 - q/2d \}$, where $\map N_q^+$ and $\map N_q^-$ are given by
\begin{equation}
    \map N_q^+(\rho) = \frac {\frac{qI}{2d} + \frac q {2d} \cdot \rho^T + (1-q)\rho} {1-\frac q 2+ \frac q {2d}} \, , \quad \map N_q^-(\rho) = \frac {I-\rho^T} {d-1} \, .
\end{equation}
    
Then we can show that both Holevo information of $\map N_q^+$ and that of $\map N_q^-$ can be achieved with the ensemble from uniform sampling of the computational basis. The Holevo information of a channel $\map C$ from system $A$ to system $B$ can be bounded as follows:
\begin{equation}
\label{eq:max_holevo}
    \chi(\map C) = \max_\xi I(M;B) = \max_\xi H(B) - H(B \mid M) \leq \log d - H^{\min}(\map C) \, .
\end{equation}
Due to the concavity of von Neumann entropy, for any input state $\rho$, we have
\begin{equation}
    H(\map N_q^+(\rho)) \geq H\left(\frac {\frac{qI}{2d} + \frac q {2d} \cdot \rho + (1-q)\rho} {1-\frac q 2+ \frac q {2d}}\right) = H\left(\frac {\frac{qI}{2d} + \frac q {2d} \cdot \rho^T + (1-q)\rho^T} {1-\frac q 2+ \frac q {2d}}\right) \, ,
\end{equation}
with equality when $\rho = \rho^T$. It follows that both $H^{\min}(\map N^+)$ and $H^{\min}(\map N^-)$ can be achieved when $\rho$ is a pure state such that $\rho = \rho^T$. 
\begin{equation}
    \begin{split}
        H^{\min}(\map N_q^+) &= \frac{-\left(\frac q d +1-q\right)\log\left(\frac q d +1-q\right) - (d-1)\frac q {2d}  \log\left(\frac q {2d}\right)}{1-\frac q 2+ \frac q {2d}} + \log \left(1-\frac q 2+ \frac q {2d} \right) \, , \\ 
        H^{\min}(\map N_q^-) &= \log (d-1) \, .
    \end{split}
\end{equation}
Finally we see that the inequality in \ref{eq:max_holevo} can be saturated for both of the channels $\map N_q^+$ and $\map N_q^-$ when the the sender's ensemble is
\begin{equation}
    p_m = \frac 1 d \, , \ \rho_m = |m\>\<m| \, , \quad m = 0, \cdots, d-1 \, .
\end{equation}
By Theorem \ref{theo:labeledrandom}, we can see that the Holevo information of the effective channel is
\begin{equation}
\label{eq:depolarising_holevo}
\begin{split}
    \chi(\map F_{ |+\>\<+|}  (\map N_q)) 
    &= \left(1 - \frac q 2 + \frac q {2d}\right) \chi(\map N_q^+) + \left(\frac q 2 - \frac q {2d}\right) \chi(\map N_q^-) \\
    &= \left(1 - \frac q 2 + \frac q {2d}\right) (\log d - H^{\min}(\map N^+)) + \left(\frac q 2 - \frac q {2d}\right) (\log d - H^{\min}(\map N^-)) \\
    &=\log d - \left(1 - \frac q 2 + \frac q {2d}\right) \log \left(1 - \frac q 2 + \frac q {2d}\right) \\
	&\quad +\left(1-q+\frac q d\right)\log\left(1-q+\frac q d\right) + (d-1)\frac q {2d}  \log\left(\frac q {2d}\right) - \left(\frac q 2 - \frac q {2d}\right) \log(d-1) \, . 
\end{split}
\end{equation}
In addition, if system $A$ has dimension 2, both $\map N_q^+$ and $\map N_q^-$ are qubit bistochastic channels which are Holevo additive \cite{king2003capacity}, then by Theorem \ref{theo:labeledrandom} the classical capacity of the effective channel $\map F_{ |+\>\<+|}  (\map N_q)$ reduces to the Holevo information given by Eq.  \ref{eq:depolarising_holevo}. 

\section{Derivation of the coherent information of the effective channel based on the depolarising channel}\label{app:coherentinfo}

In this section, we derive the coherent information of the effective channel $\map F_{ |+\>\<+|}  (\map N_q)$ when the state $I/d$ is its input, which is equal to the coherent information of the state shared by Alice and Bob after Alice prepares a maximally entangled state $|\Phi\>$ and send half of it to Bob via the effective channel:
\begin{equation}
    I_c(I/d, \map F_{ |+\>\<+|}  (\map N_q)) = H \left( \map F_{ |+\>\<+|}  (\map N_q)(I/d) \right) - H \left( \map (\map F_{|+\>\<+|}(\map N_q) \otimes \map I)(|\Phi\>\<\Phi|) \right) \, ,
\end{equation}
where $\map F_{ |+\>\<+|}  (\map N_q)(I/d)$ is given by
\begin{equation}
    \map F_{ |+\>\<+|}  (\map N_q)(I/d) = \left(\frac 1 d - \frac q {2d} + \frac q {2d^2} \right) (I \otimes |+\>\<+|) + \left( \frac q {2d} - \frac q {2d^2} \right) (I \otimes |-\>\<-|) \, ,
\end{equation}
and its entropy is
\begin{equation}
\label{eq:entropy1}
H \left( \map F_{ |+\>\<+|}  (\map N_q)(I/d) \right) = 
    -d \left(\frac 1 d - \frac q {2d} + \frac q {2d^2} \right) \log \left(\frac 1 d - \frac q {2d} + \frac q {2d^2} \right) - d \left( \frac q {2d} - \frac q {2d^2} \right) \log \left( \frac q {2d} - \frac q {2d^2} \right) \, ,
\end{equation}
and $(\map F_{ |+\>\<+|}(\map N_q) \otimes \map I)(|\Phi\>\<\Phi|)$ is given by
\begin{align}
    (\map F_{ |+\>\<+|}(\map N_q) \otimes \map I)(|\Phi\>\<\Phi|)
    &= \frac{q}{2d^2}(I \otimes (|0\>\<0|+|1\>\<1|) + {\tt SWAP} \otimes (|0\>\<1|+|1\>\<0|)) + (1-q)|\Phi\>\<\Phi| \otimes |+\>\<+| \\
    &= \left[\frac{q}{2d^2}(I + {\tt SWAP}) + (1-q)|\Phi\>\<\Phi| \right] \otimes |+\>\<+| + \left[\frac{q}{2d^2}(I - {\tt SWAP})\right] \otimes |-\>\<-| \, ,
\end{align}
and its entropy is
\begin{align}
\label{eq:entropy2}
    H \left( \map F_{ |+\>\<+|}  (\map N_q) \otimes \map I)(|\Phi\>\<\Phi|) \right)
    &= -(d^2-1) \left(\frac{q}{d^2}\right) \log \left(\frac{q}{d^2}\right) - \left(1-q+\frac{q}{d^2}\right) \log \left(1-q+\frac{q}{d^2}\right) \, .
\end{align}

Combining Eq. \ref{eq:entropy1} and Eq. \ref{eq:entropy2}, we obtain the coherent information:
\begin{multline}
    I_c(I/d, \map F_{ |+\>\<+|}  (\map N_q)) = 
    -d \left(\frac 1 d - \frac q {2d} + \frac q {2d^2} \right) \log \left(\frac 1 d - \frac q {2d} + \frac q {2d^2} \right) \\
     - d \left( \frac q {2d} - \frac q {2d^2} \right) \log \left( \frac q {2d} - \frac q {2d^2} \right) +
    (d^2-1) \left(\frac{q}{d^2}\right) \log \left(\frac{q}{d^2}\right) + \left(1-q+\frac{q}{d^2}\right) \log \left(1-q+\frac{q}{d^2}\right) \, .
\end{multline}

\end{document}